# LANNDD - A line of liquid argon TPC detectors scalable in mass from 200 Tons to 100 KTons

David B. Cline[a], Fabrizio Raffaelli[b] and Franco Sergiampietri[a,b]

[a] *Department of Physics, UCLA, Los Angeles, CA 90024, USA*
[b] *INFN - Sezione di Pisa, Largo B. Pontecorvo 3, 56127 PISA, Italy*

## Abstract

We propose a scalable line of liquid argon TPC detectors based on a three dimensional cubic frame array immersed on a common liquid argon volume. The paper describes general lines, main construction criteria, crucial points, parameters and required preliminary R&D activities for the construction of detectors with active mass ranging from 200 ton to 100 kTon. Such detectors appear as unique for supernova detection, proton decay, LBL neutrino physics and other astropaticle physics applications.

## 1 Introduction

This paper takes up the general considerations already described in a previous work [1], but a different solution is worked out to enhance the scalability and the matching of the shape with the intrinsic geometry of the time-projection imaging mechanism.

The on-surface test of the 300-ton ICARUS module made in 2001 [2], demonstrated the 3D-imaging potentiality of this detection technique, only comparable to that of a bubble chamber but extended at many order of magnitude larger volumes. Due to its multi-module configuration and to its large liquid nitrogen consumption (~1 liquid $m^3$/hour), the 300-ton geometry and construction technique appears unlikely suitable for the design of a much larger mass detector.

The preliminary study [3],[4] on the modularity and the shape for a future large-scale detector, named LANNDD (Liquid Argon Neutrino and Nucleon Decay Detector), gave indication on the advantages of a single-module cryostat. This conclusion is valid only if valid solutions are found for keeping the same construction/operation quality required for a detector based on an ultra high purity (UHP) liquefied noble gas and for coping with the engineering and safety issues related to the extended scale.

The huge costs, the multi-year construction and commissioning phase required for such a project can only be justified by a credible physics plan that should include natural and artificial neutrino oscillation physics and, seen the unique large instrumented mass combined with the low energy detection threshold, nucleon decay physics. This programme can only be performed by conceiving the detector as sited in an underground laboratory (in the USA, in Europe or in Japan) and in the beam line (in-axis or off-axis) of existing or future neutrino beams.

In the following we describe a further step on the study of possible realistic solutions and the required preliminary R&D activities for such a detector. We would like to underline that, in a large scale, the main difficulties do not arise from the detection technique, that even with some possible improvements or changes is the well-established ICARUS one, but from the engineering choices in the cryostat configuration affecting performances (purity levels, thermal insulation), mechanical stability, safety and the related construction and operation costs.

The design for a full scale detector become credible through the project and construction of a reduced scale prototype, configured as scalable to the final mass. Due to its costs, such a prototype should be able to provide information not only on the optimized construction technique but also on its detection performances with cosmic particles and as near detector in a neutrino beam. A useful utilization is also its calibration for reconstructed energy, momentum and position in a hadronic and electron test beam. A detector with an active mass in the 100 - 200 Ton range seems the right choice for such purpose.

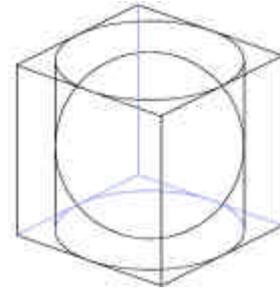

**Figure 1**

## 2   Project guidelines

In designing a line of mass scalable detectors we referred to guidelines and construction criteria described below.

*Modularity* - The preliminary study described in Ref. [3] pointed out the advantages in fiducial volume, number of channels, heat input, required electric power of a single-module configuration compared to a multi-modular one.

*Shapes* - Fiducial-to-active volume ratio, heat input, wall out-gassing, number of wires and electronic channels are related to the surface-to-volume ratio (S/V).

A cubic vessel has the same S/V ratio than a cylindrical or spherical one inscribed in it (Fig. 1), but with volume ratios

$$V_{cube} : V_{cylinder} : V_{sphere} = 1 : 0.5 : 0.8$$

The spherical shape has in principle the highest advantages for the mechanical stiffness but, in practice, can be considered only with a radius equal to the maximum drift path (only one equatorial, bi-face wire chamber)

The cylindrical shape appears a good compromise between time projection imaging mechanism and required structural stiffness, but with a non optimal ratio between the instrumented LAr volume and the total LAr volume if the wire chamber are parallel to the cylinder axis or with a non-optimal distribution of the wire lengths (resulting in a larger number of channels) if the chambers have a circular or polygonal shape in planes orthogonal to the cylinder axis.

The cubic (parallelepiped) shape is the most adapt for the projection imaging geometry and for the instrumented-to-total LAr volume ratio, but requires special care for the mechanical stiffness in the design of the bearing structure.

*Construction costs* – Having chosen a single module configuration, as first approximation, cryostat, liquid argon and purification system have costs roughly independent from the shape of the detector. Other construction costs depend mainly on the number of wires and associated electronic channels and can be decreased by:

 a) reducing the wire chamber number by widening the drift region between each cathode and the facing wire chamber (drift paths ? 5 meters). Operating with such drift lengths implies the use of high voltages in the range 200~300 kV and sets serious constraints on the LAr purity (impurities = 10~50 ppt $O_2$ equivalent).

 b) developing a new generation electronics:
   - higher S/N ratio, to balance the long drift attenuation
   - medium scale integration (goal: 2~10 times cheaper)
   - safe operation in LAr, with front-end directly connected to the wires to avoid the capacitance load of long signal cables
   - amplified signals multiplexed in LAr, to decrease the number of signal feedthroughs, the heat input and the argon contamination by cables

 c) two read-out wire planes per chamber (orientation 0° and 90°) with all wires with equal length.

Due to the previously mentioned project-implied conditions (huge costs, multi-year construction and commissioning), we cannot, nevertheless, economize in any crucial points and understand the mistakes at the moment of first putting in operation. In order to reach and maintain during years the required level of purity we consider as inalienable the following construction criteria and conditions:

  - Possibility of generating vacuum inside the inner vessel and of checking its tightness.

  - Wise choice of construction materials: use of stainless steel for the inner vessel and for cathodes, wire chamber frames and electrical field shaping electrodes; (a possible use for the outer vessel of alternative and cheaper alloys, for instance as CORTEN, should be evaluated).

  - Continuous, adiabatic argon purification in liquid phase.

  - UHP and UHV standards for any device and cryogenic detail (flanges, valves, pipes, welding) in contact with the argon.

*Running costs* – Running costs are mainly related to the efficiency of the thermal insulation and to the number of channels. Vacuum insulation, joint to the use of superinsulation jacket around the cold vessel, should be considered as the primary choice.

An optimized thermal insulation, with low evaporation rate for LAr and low electric power involvement for cryogenerators or re-condensers, is also a must to smoothly and safely operate the detector underground during tens of years.

## 3 The LANNDD cellular structure

We have studied a solution that allows a continuous (not segmented) active LAr volume (high fiducial volume) contained in a cryostat based on a multi-cell mechanical structure. The main idea is not to scale in size a single cell, but to reach the required volume by extending the number of equal cells (see Figure 2a and 2b). This cellular geometry concerns only the mechanical structure while the active LAr drift volumes and read-out chambers are continuous across each 2D cell array (see Figure 3 and Figure 6b).

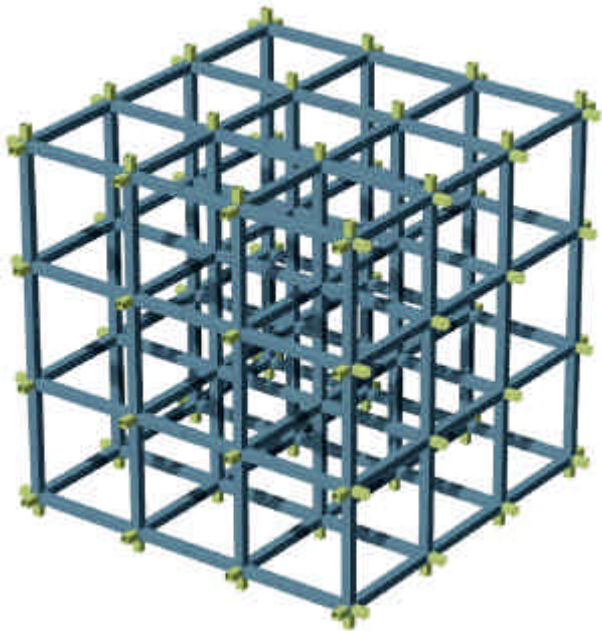
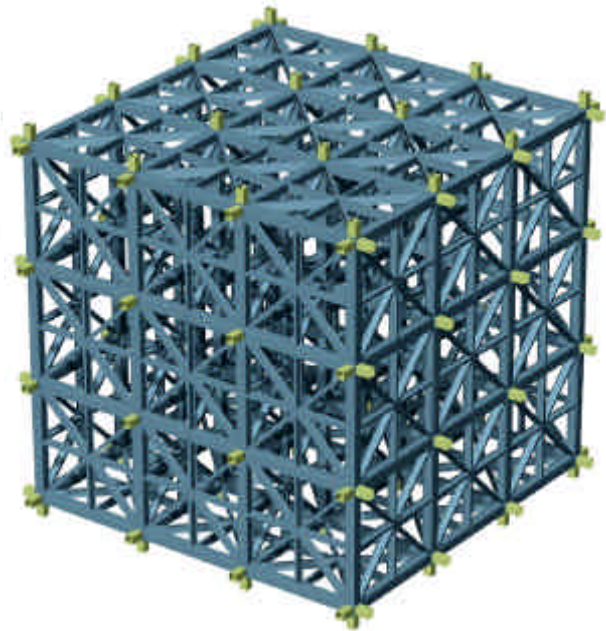

**Figure 2a** – Simplified mechanical structure for the case of $3^3$ cell array.

**Figure 2b** – Inner cell array structure with reinforced cell sides.

The inner cellular structure is a repetitive 3D array of cubic boxes (Fig. 2a) with outer faces reinforced by double cross beams (Fig. 2b, a similar reinforcement is adopted also for the inner cell faces orthogonal to the electron drift direction). The connection to the outer (warm) wall is obtained by thermal bridge beams, elastically linked at middle length of the cold beams.

Peripherical sides of the 3D cell array are outfitted with double-layer walls to form a cubic cold box. This solution allows a cubic shape composed by $n^3$ cells, $5m \times 5m \times 5m$ in size each.

The cryostat is then made by an inner and an outer cubic box elastically linked in between by thermal bridge beams. The thermal bridge beams are allowed to slide with respect to a central point in the basement and in the other outer warm faces to compensate the thermal shirking of the inner box.

Both the inner and outer boxes have double-face linked walls, to increase their stiffness (see below) and to allow a sector by sector vacuum tightness check.

The lattice beam structure is vacuum-tight as well and its beam structure is used for a volume distributed $LN_2$ cooling. $LN_2$ flow and temperature can be controlled independently in each 2D cell layer at different heights to limit temperature gradients and related vortexes in the LAr (dangerous for the induced microphonic noise in the wires).

Vacuum tightness in the beam structure and inside the double layer walls provides a way to check the full tightness of the inner and outer boxes.

The inner box is fully wrapped by superinsulation layers.

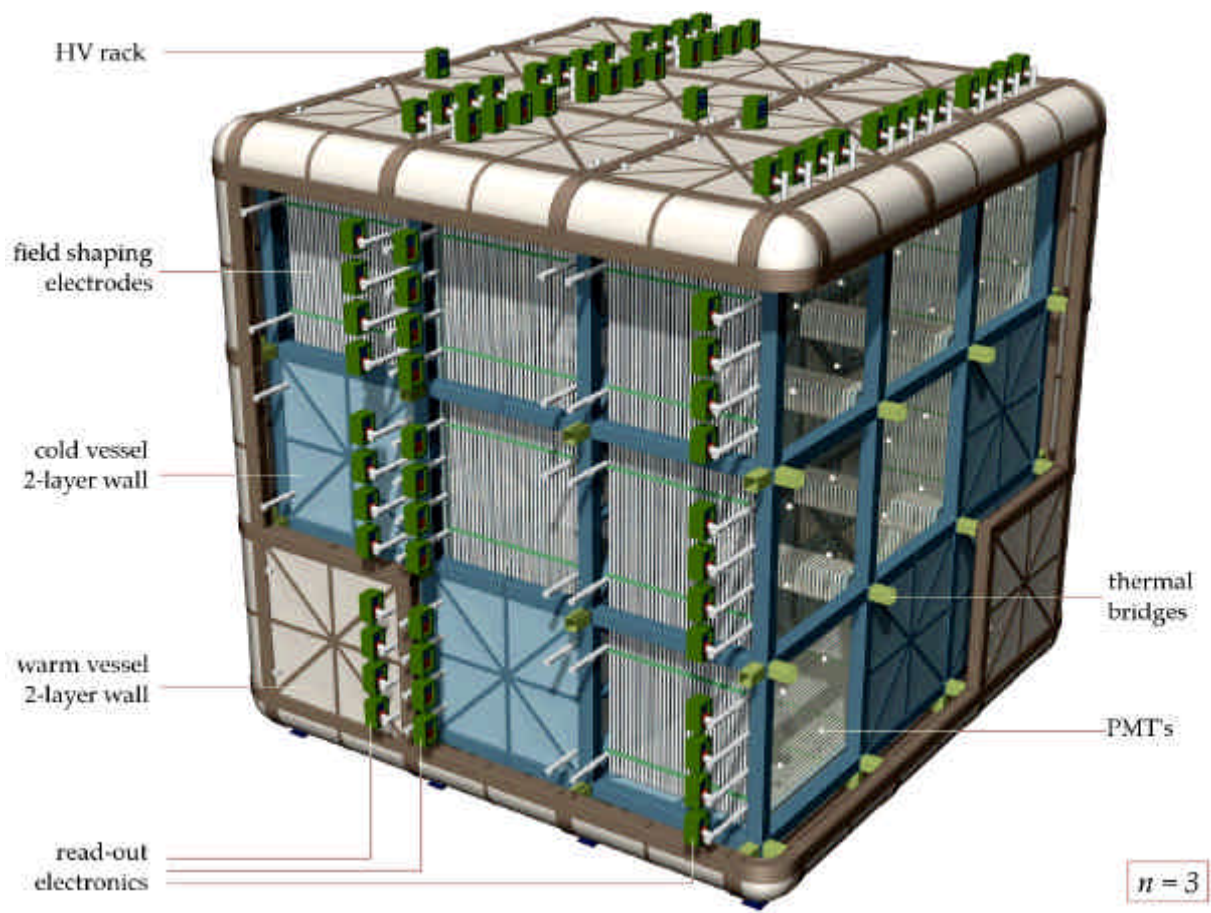

**Figure 3** - Cutaway view of the proposed cryostat design. The $3^3$ configuration is shown as example.

# 4 Mechanical consideration and finite element analysis

The design of large rectangular vessel requires special attentions:

1) To obtain the right bending stiffness of the plate without using a large quantity of material
2) To control the large moment in the corners of the box. The size of the cell must be able to avoid instability problems in the compressed part of the plate.
3) The welding connection must be able to transfer the in between shear and the contribution to the deflection caused by it must be managed.

For the outer faces, outfitted with double layer (plate) walls the increase of the bending stiffness can be achieved by spacing the two plates:

In the process of spacing the two plates are one in traction and the other in compression. This makes the stress uniform in each plate. The shear can contribute to the overall deflection. We need to control the instability regions of the structure where there are compressive stresses.

A model was done for the $3^3$ cell configuration using stainless steel beams - 500 × 500 *mm* in size, 20 *mm* thick - and wall panels made by two linked plates (thickness = 8 *mm*, distance = 484 *mm*, total wall thickness = 500 *mm*, equivalent to a 200 *mm* thick solid plate, see Figure 4).

The single cell inside size is 5 *m* and G10 blocks are supposed to space inner and outer tanks as thermal bridges.

The pressure is 0.1 *Mpa*, from outside, on the outer tank and 0.3 *Mpa*, from inside, on the inner tank.

Additional considerations need to be addressed concerning the availability of the size material, process and costs.

However there is a lot possibility to carry out an optimization of the structure to guarantee the mechanical strength and stability.

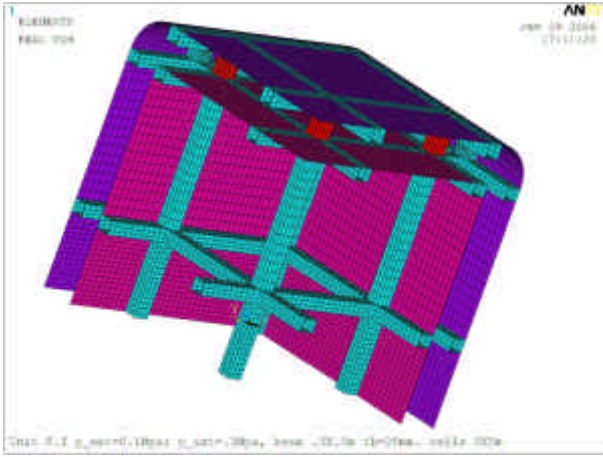
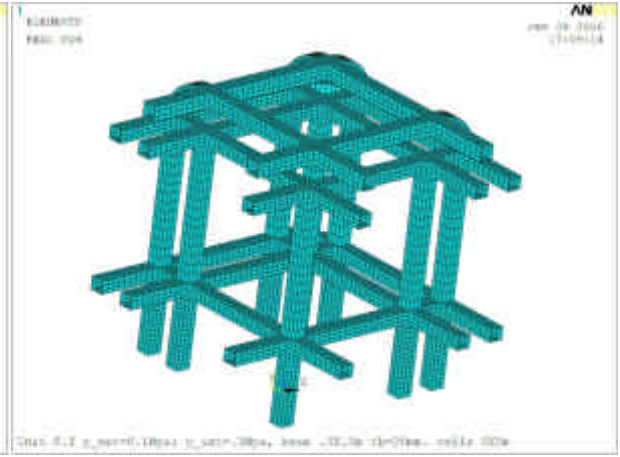

**Figure 4a** - Due to the cubic symmetry, FEA can be performed on a model representing only 1/8 of the 3×3×3 cell detector by applying proper symmetry boundary conditions

**Figure 4b** – Beam structure

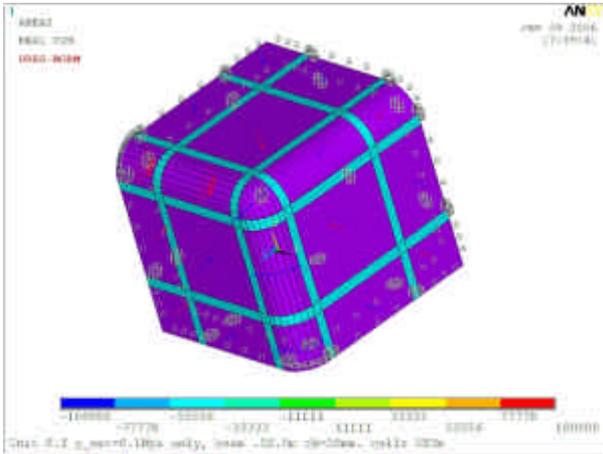
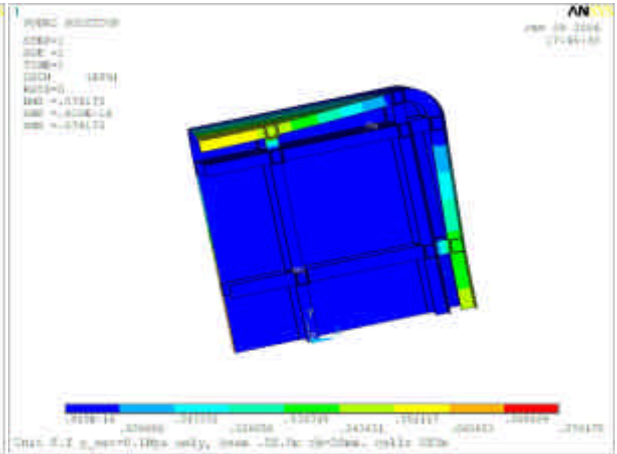

**Figure 4c** - Symmetry boundary conditions

**Figure 4d** - Case with only external pressure (vacuum inside). The units are in *meter Pa*. The max displacement is *?z* = 7.8 *mm*.

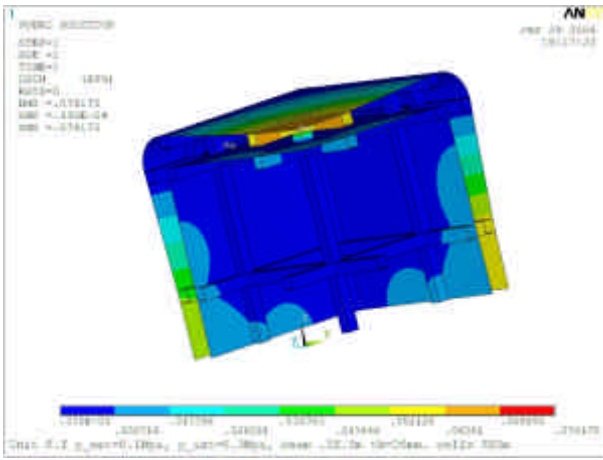
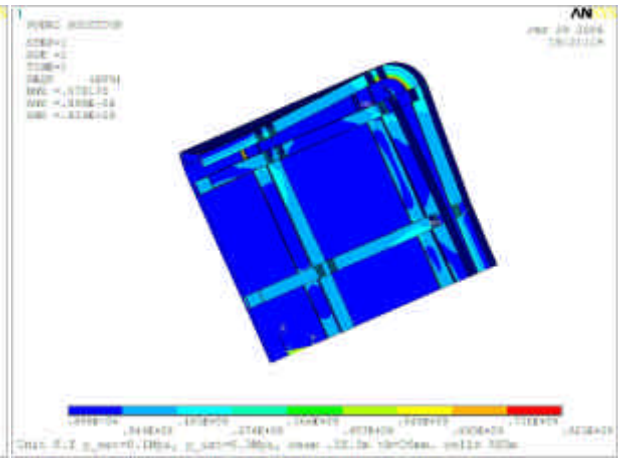

**Figure 4e** - Internal and external pressure and vacuum between the two layers give the same results for *?z* than Fig. 5d.

**Figure 4f** - This picture shows the stress on the corners.

# 5   LANNDD - 001

A single cell detector is proposed (Figure 5) as study prototype to analyze the mechanical stiffness of the inbox-outbox complex at room and at low temperature and in condition to simulate the operating conditions for a multi-cell configuration[a].

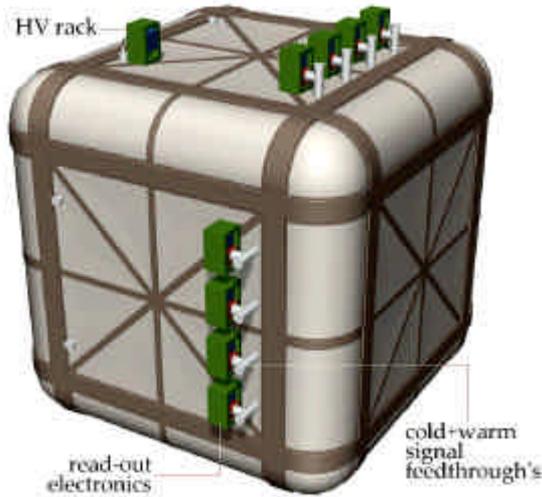

**Figure 5a** – Outside view of the single cell configuration.

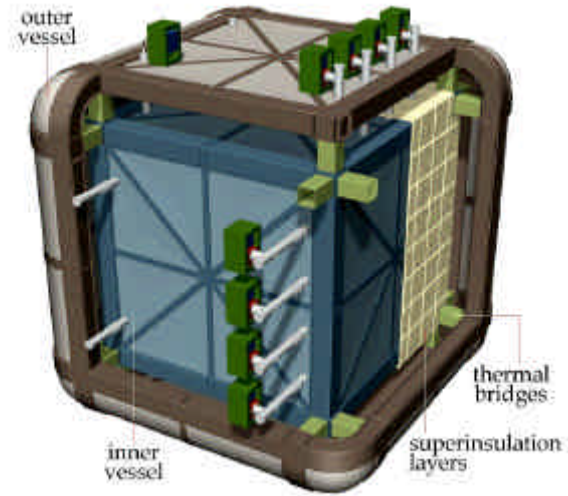

**Figure 5b** – Cutaway view of the single cell detector showing the inner and outer vessels, the superinsulation layer and the thermal bridge spacers.

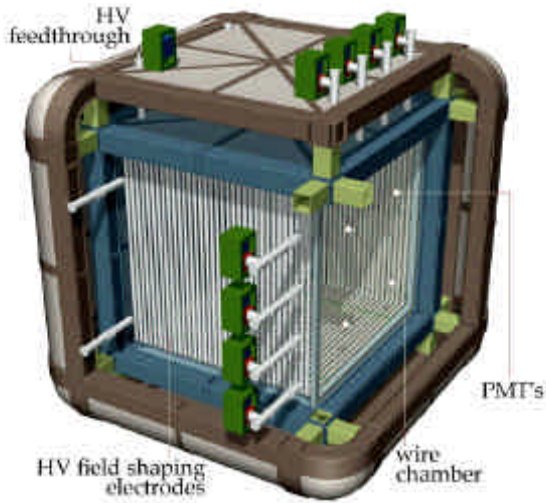

**Figure 5c** - Cutaway view of the single cell detector showing the inner detector arranged inside the inner vessel.

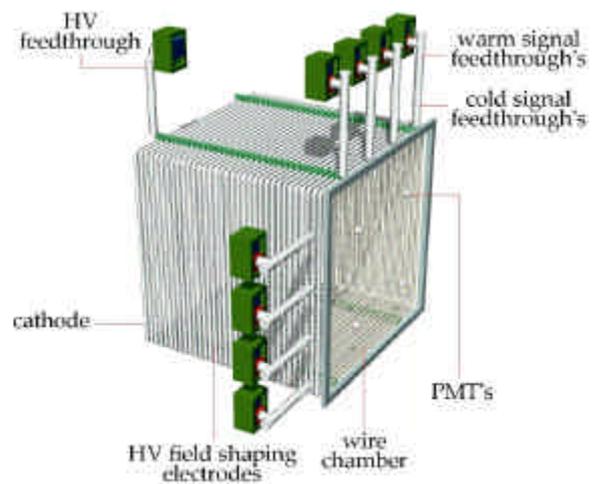

**Figure 5d** – Single cell inner detector showing the HV field shaping cage, the cathode, the wire chamber and the PMT's.

---

[a] Maximum pressure for the $8^3$ cell configuration = 6.6 *bar* (47.5 *m* LAr maximum height) + 1 *bar* (insulation vacuum).

The detector is useful for developing and testing HV system, readout electronics, purification and cooling systems, acquisition and reconstruction software and all repetitive details and solutions to adopt for the full scale detector. Such a prototype is suitable for providing information on detection performances with cosmic particles and as near detector in a neutrino beam as well as on energy/momentum/position/linearity calibration in a hadronic and electron test beam.

The instrumented LAr has a volume of 125 $m^3$ and an instrumented mass of 175 *Ton*.

# 6   LANNDD - 027

An intermediate mass detector to be used as middle distance detector or in an off-axis neutrino beam line is made by $3^3$ = 27 basic cubic cells (see Figure 6). The active volume is split into 3 drift volumes 17×17×5 $m^3$, with two 0.9 *m* thick layers in between to host the inner beam structure lattice and photo-multiplier arrays.

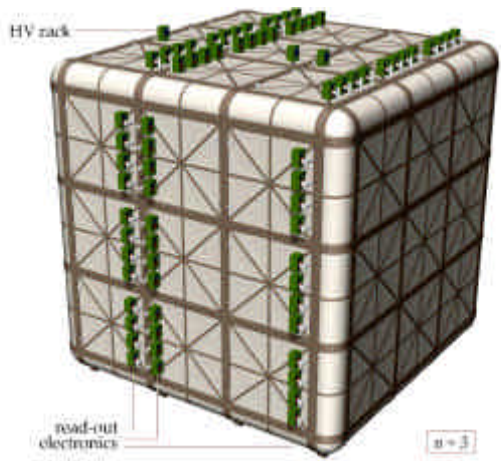
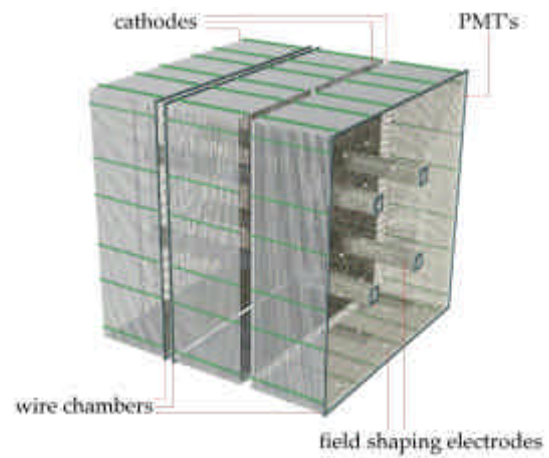

**Figure 6a** - Outside view of the 27-cell configuration.

**Figure 6b** – 27-cell inner detector showing the HV field shaping cage, the cathode, the wire chamber and the PMT's. The arrangement inside the cryostat is shown in Figure 3.

The instrumented LAr has a volume of 4'400 $m^3$ and an instrumented mass of 6.1 *KTon*.

# 7   LANNDD - 512

The maximum mass detector for use as far detector in a LBL ?-beam and for nucleon decay studies in an underground site is made by $8^3$ = 512 basic cubic cells (Figure 7).

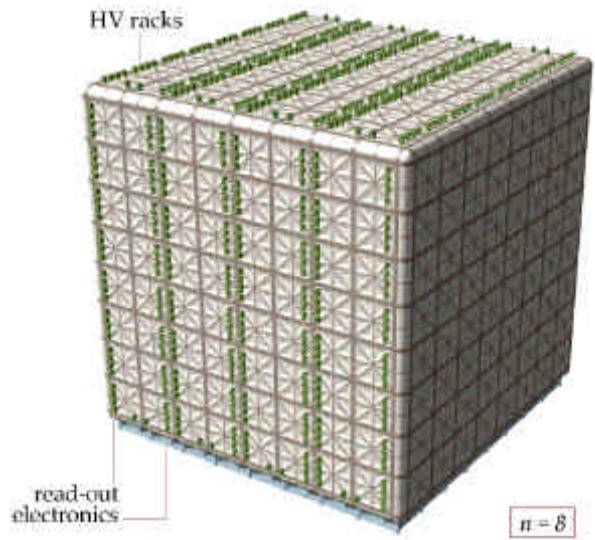 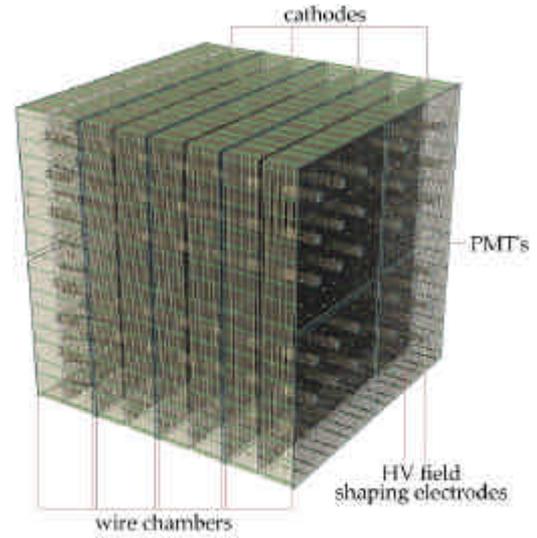

**Figure 7a** - Outside view of the 512-cell configuration.

**Figure 7b** - 512-cell inner detector showing the HV field shaping cage, the cathode, the wire chamber and the PMT's.

The instrumented LAr has a volume of 85'280 $m^3$ and an instrumented mass of 119 *KTon*.

# 8   Scaling

The proposed configuration allows and easy way to scale the detector volume to a value optimized in terms of physics requirements, mechanical stability and safety issues and costs by choosing the proper number of cells. The final aspects of the three considered configurations are show together in Figure 8 for comparison. Their main parameters are listed in Table 1 where some other configuration is included.

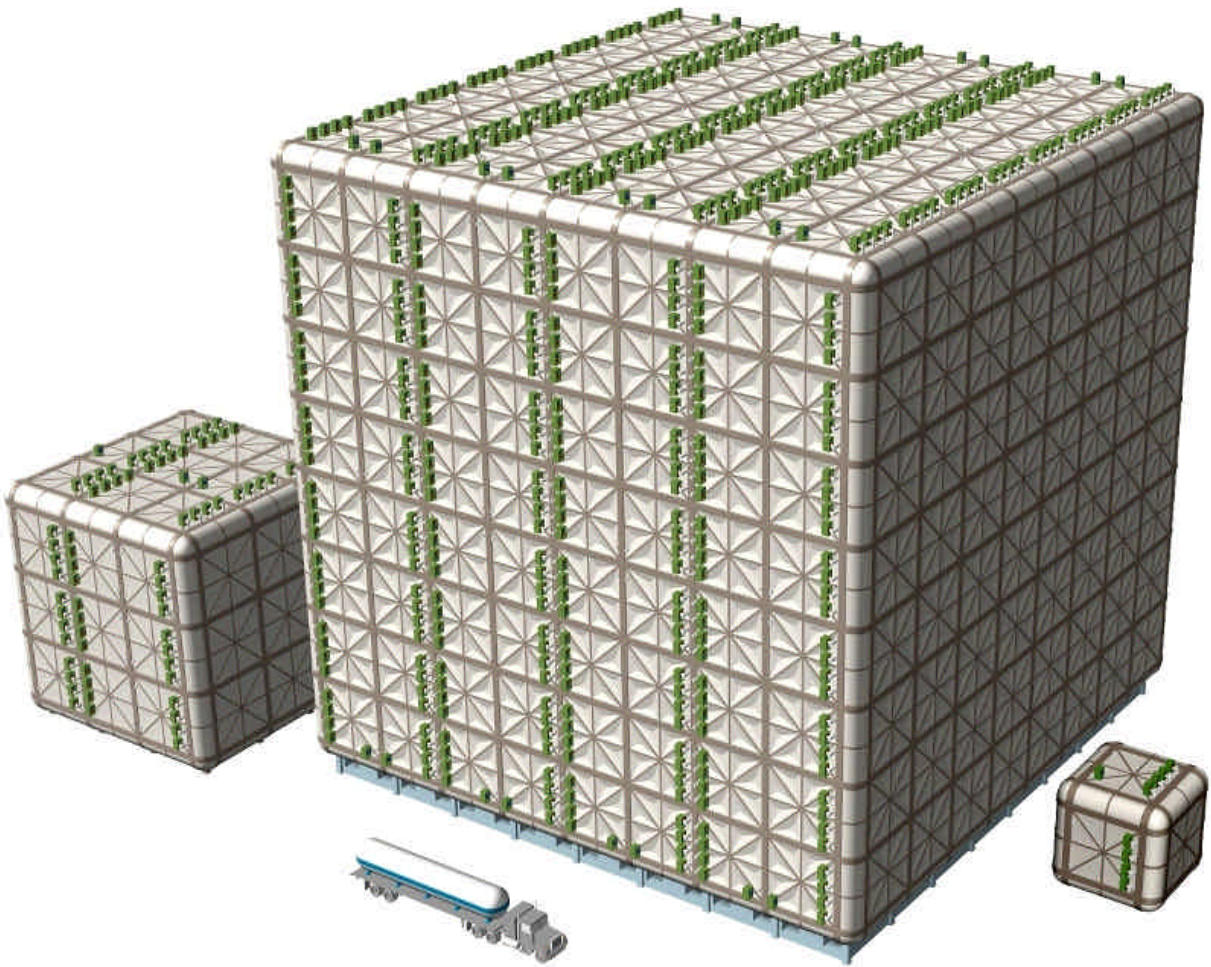

**Figure 8** – The single- cell (right), the 27-cell (left) and the 512-cell (middle) detectors are shown together for a size comparison.

Table 1 – Main parameters for detector configuration with 1 – 512 cubic cells

| | | | | | | |
|---|---|---|---|---|---|---|
| Active LAr mass, *Ton* | 184.4 | $1.7 \times 10^3$ | $6.1 \times 10^3$ | $14.8 \times 10^3$ | $48.9 \times 10^3$ | $118.5 \times 10^3$ |
| Number of cells/side | 1 | 2 | 3 | 4 | 6 | 8 |
| Total number of cubic cells | 1 | 8 | 27 | 64 | 216 | 512 |
| Total LAr volume, $m^3$ | 166.4 | $1.5 \times 10^3$ | $5.3 \times 10^3$ | $12.8 \times 10^3$ | $44.1 \times 10^3$ | $105.5 \times 10^3$ |
| Total-to-active volume ratio | 1.25 | 1.22 | 1.21 | 1.20 | 1.25 | 1.24 |
| Total heat input, *KW* | 0.3 | 1.0 | 2.1 | 3.7 | 8.1 | 14 |
|    Equiv. $LN_2$ consumption, $m^3/d$ | 0.18 | 0.63 | 1.36 | 2.36 | 5.2 | 9.1 |
|    Equiv. el. power for cooling, *KW* | 2.8 | 9.8 | 21 | 37 | 81 | 140 |
| Number of wire chambers | 1 | 2 | 3 | 4 | 24 | 32 |
| Wire length, *m* | 5.1 | 11.1 | 17.1 | 23.1 | 17.1 | 23.1 |
| Number of channels | $3.3 \times 10^3$ | $14 \times 10^3$ | $33 \times 10^3$ | $59 \times 10^3$ | $270 \times 10^3$ | $483 \times 10^3$ |
| El. Power for electronics, *KW* | 4 | 16 | 38 | 68 | 310 | 555 |

# 9  Conclusions

The cubic shape optimizes the ratio between active and total LAr volumes at values 80-83% and allows a detector made by equal length (and impedance) wires. This shape minimizes the surface-to-volume ratio as well, with relevant advantages in the thermal insulation and in the reduction of wall out-gassing.

The proposed cellular structure combines mechanical stiffness with large, continuous active LAr volumes. Finite element analysis gives satisfactory and promising indications.

The residual non-active LAr layers due to the inner beam structure are fruitfully used for:

 a) distributed $LN_2$ cooling
 b) positioning of a 3D array of photomultipliers with a complete coverage of the active LAr.

The test and optimization of this solution (and of all the concerned techniques) can be performed on a single-cell prototype.

Vacuum insulation, wise choices for construction materials, UHP & UHV standards for the components assure an exceptionally cheap and safe operation for long term running and widely compensate their construction costs.

We have tried to underline the main guidelines and crucial points for the design of a multi-kTon LAr imaging detector. The extremely low $LN_2$ consumption, due to the double-wall vacuum insulation, and the reduced number of channels, due to the monolithic detector geometry and to the long drift, make the described example solution worthy of in-deep working out, in an engineering frame, for a realistic single cell, 170 Ton prototype. The structure of this detector-cryostat complex is such to allow scaling it up to the required greater volumes by choosing the proper cell multiplicity.

The described analysis suggests R&D activities on long path drift, on cryogenic adiabatic pumps, on cold signal feedthroughs and on a new medium scale integrated cryogenic electronic chain. Such crucial items are the main programme of the R&D activity of Ref. [5].

# References

[1]  D.B. Cline and F. Sergiampietri, *"A Concept for a Scalable 2 kTon Liquid Argon TPC Detector for Astroparticle Physics"*, http://arxiv.org/abs/astro-ph/0509410

[2]  ICARUS Collaboration, *"Design, construction and tests of the ICARUS T600 detector"*, Nuclear Instruments and Methods-Section A, **527** (2004) 329-410

[3]  F. Sergiampietri, *"On the possibility to extrapolate Liquid Argon Technology to a super massive detector for a future Neutrino Factory"*, talk given at the 3rd INTERNATIONAL WORKSHOP ON NEUTRINO FACTORY BASED ON MUON STORAGE RINGS (NuFACT'01), Tsukuba-Japan, May 2001

[4]  D.B. Cline, F. Sergiampietri, J. G. Learned, K. McDonald, *"LANNDD: a massive liquid argon detector for proton decay, supernova and solar neutrino studies, and a neutrino factory detector"*, Proceedings of the 3rd International Workshop on Neutrino Factory based on Muon Storage Rings (NuFACT'01), Tsukuba-Japan, May 2001, Nuclear Instruments and Methods-Section A, **503** (2003) 136

[5]  A. Bueno, C. Cerri, D. B. Cline, S. Navas, R. Pazzi, F. Sergiampietri, X. Yang, H. Wang, *"LANNDD-5mD R&D PROPOSAL"*, CERN-SPSC-2004-033, SPSC-I-230, 30/09/2004